\documentclass[11pt]{article}
\usepackage{geometry}                
\usepackage{graphicx}
\usepackage{amssymb, amsmath}
\usepackage{epstopdf}
\usepackage{url}
\usepackage{subfigure}
\begin{document}

\begin{centering}

{\LARGE A Bayesian Approach To Histogram Comparison}

\hspace{5mm}

Michael Betancourt \footnote{betan@mit.edu}

\textit{Massachusetts Institute of Technology, Cambridge, MA 02139}

\end{centering}

\hspace{3mm}

\begin{abstract}

Determining if two histograms are consistent, whether they have been drawn from the same underlying distribution or not, is a common problem in physics.  Existing approaches are not only limited in power but also inapplicable to histograms filled with importance weights, a common feature of Monte Carlo simulations.  From a Bayesian perspective, the comparison between a single underlying distribution and two underlying distributions is readily solved within the context of model comparison.  I introduce an implementation of Bayesian model comparison to the problem, including the extension to importance sampling.

\end{abstract}

\section{Histogram Comparison}

The histogram is a fundamental tool in physics, providing a powerful yet accessible means of non-parameteric inference.  Often, however, analyses are concerned not with the underlying distribution itself but rather the consistency between multiple histograms.  Comparing different analysis techniques, for example, requires the comparison of each respective analysis output given the same input data.  Perhaps the most ubiquitous example is the validation of Monte Carlo simulations against data.  

Although numerous approaches to the problem exist, none are without their limitations.

\section{Orthodox Approaches}

\subsection{The Komolgorov-Smirnov Test}

When comparing the consistency of two one-dimensional samples, $\mathbf{y}_{1}$ and $\mathbf{y}_{2}$, standard frequentist methodologies turn to significance testing, in particular the Kolmogorov-Smirnov (KS) test \cite{Papoulis2002}.  Here the null hypothesis that the two samples are consistent with draws from the same distribution is accepted or rejected based on the sampling distribution of the test statistic

\begin{equation*}
q = \max_{x} \left| F_{\mathbf{y}_{1}} \left( x \right) - F_{\mathbf{y}_{2}} \left( x \right) \right|,
\end{equation*}

\pagebreak

\noindent where $F_{\mathbf{z}} \left( x \right)$ is the empirical distribution of the sample $\mathbf{z}$,

\begin{equation*}
F_{\mathbf{z}} \left( x \right) = \sum_{i = 1}^{n} I \left( x_{i} \right),
\end{equation*}

\noindent with $I \left( x_{i} \right)$ the indicator function

\begin{equation*}
I \left( x_{i} \right) = \left\{ \begin{array}{rc} 0,& x < x_{i} \\ 1,& x \geq x_{i} \end{array} \right. .
\end{equation*}

In practical applications, the sampling distribution for $q$ is usually approximated with a large $n$ limit such that the KS test becomes extremely conservative (in more technical terms, the Type II error of the test is small only when the samples are large).  Moreover, the sampling distribution of the weighted samples from importance sampling cannot be approximated so easily and the proper application of the KS test to simulated events is infeasible.

Given histograms instead of the individual samples, the empirical distributions must be approximated by the bin contents themselves and the test statistic $q$ reduces to the maximum difference between bins.  Information in all other bins is ignored, no matter now relevant it might be.  Considering also the usual faults of frequentist significance tests \cite{Bernardo2000, Jaynes2003, MacKay2003}, the application of the KS test to the problem of histogram comparison leaves much to be desired.

\subsection{Other Approaches}

Various procedures have also been developed in various communities.  One approach subtracts one histogram from the other before performing a significance test on the hypothesis of a constant residual across all bins.  The sampling distribution of the residuals, however, does not admit familiar tests such as $\chi^{2}$ outside of the limit of large bin contents.  Still, the limit is often assumed and the resulting $\chi^{2}$ taken as a measure of consistency between the two histograms.

A similar approach utilizing the quotient of the two histograms fares even worse.  The usual linearized Gaussian approximations to the resulting bin quotients are not very accurate, and the subsequent $\chi^{2}$ test is even more misleading than for the bin residuals.  More accurate considerations \cite{Hinkley1969} become considerably more involved, and significance testing much more challenging.

Both approaches are further limited by the dependence on frequentist significance testing.

\section{The Bayesian Perspective}

From the Bayesian perspective, the problem of histogram comparison becomes one of model comparison between $\mathcal{S}$, the model where the two histograms are drawn from the same distribution, and $\bar{ \mathcal{S} }$, the model where they are drawn from distinct distributions.  The probability of $\mathcal{S}$ follows from an application of Bayes' Theorem on the model evidences,  $p \left( \mathbf{m}, \mathbf{n} | \mathcal{S} \right)$ and $p \left( \mathbf{n}, \mathbf{m} | \bar{\mathcal{S}} \right)$,

\begin{equation*}
p \left( \mathcal{S} | \mathbf{m}, \mathbf{n} \right) = \frac{ p \left( \mathbf{m}, \mathbf{n} | \mathcal{S} \right) p \left( \mathcal{S} \right) }{ p \left( \mathbf{m}, \mathbf{n} | \mathcal{S} \right) p \left( \mathcal{S} \right)+ p \left( \mathbf{m}, \mathbf{n} | \bar{ \mathcal{S} } \right) p \left( \bar{ \mathcal{S} } \right) },
\end{equation*}

\noindent where $\mathbf{m}$ and $\mathbf{n}$ are the bin populations of the two histograms.  In practice, a uniform prior is taken and the probability reduces to

\begin{equation*}
p \left( \mathcal{S} | \mathbf{m}, \mathbf{n} \right) = \frac{ p \left( \mathbf{m}, \mathbf{n} | \mathcal{S} \right) }{ p \left( \mathbf{m}, \mathbf{n} | \mathcal{S} \right) + p \left( \mathbf{m}, \mathbf{n} | \bar{ \mathcal{S} } \right)  }.
\end{equation*}

Subsequently, $\mathcal{S}$ best models the data when $p \left( \mathcal{S} | \mathbf{m}, \mathbf{n} \right) > 0.5$ while $\bar{\mathcal{S}}$ is superior when $p \left( \mathcal{S} | \mathbf{m}, \mathbf{n} \right) < 0.5$.  When $p \left( \mathcal{S} | \mathbf{m}, \mathbf{n} \right) = 0.5$, however, the interpretation is more subtle.  The intermediate case can arise not only when each bin is independently ignorant, but also when individual bins are best described by different models and the discrepancies negate the respective contributions to the model evidences.

In order to discriminate between ignorance and disagreement, first note that the underlying multinomial distributions modeling the histogram bin contents are well described by independent Poisson distributions.  The resulting evidences conveniently factor,

\begin{equation*}
p \left( \mathbf{m}, \mathbf{n} | \mathcal{M} \right) = \prod_{i = 1}^{N} p \left( m_{i}, n_{i} | \mathcal{M} \right),
\end{equation*}

\noindent which enables an explicit mixture model,

\begin{align*}
p \left( \mathbf{m}, \mathbf{n} | \pi \right) &= \prod_{i = 1}^{N} p \left( m_{i}, n_{i} | \pi \right) \\
p \left( \mathbf{m}, \mathbf{n} | \pi \right) &= \prod_{i = 1}^{N} \left[ \pi p \left( m_{i}, n_{i} | \mathcal{S} \right) + \left(1 - \pi \right) p \left( m_{i}, n_{i} | \bar{ \mathcal{S} } \right) \right] ,\\
\end{align*}

\noindent with the mixture posterior

\begin{equation*}
p \left( \pi | \mathbf{m}, \mathbf{n} \right) \propto p \left( \mathbf{m}, \mathbf{n} | \pi \right) p \left( \pi \right).
\end{equation*}

\begin{figure}
\centering
\subfigure[]{\includegraphics[width=1.9in]{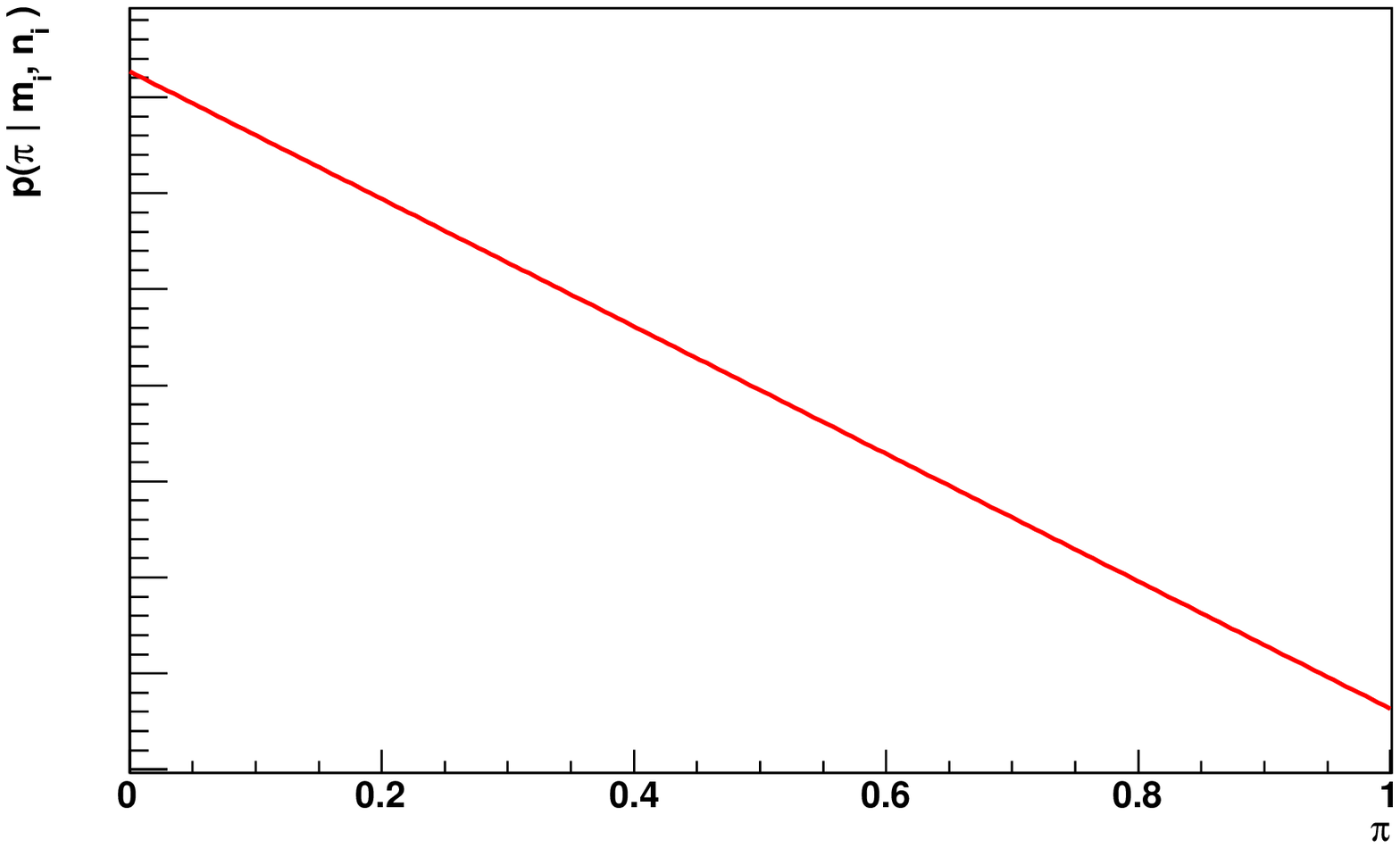}}
\subfigure[]{\includegraphics[width=1.9in]{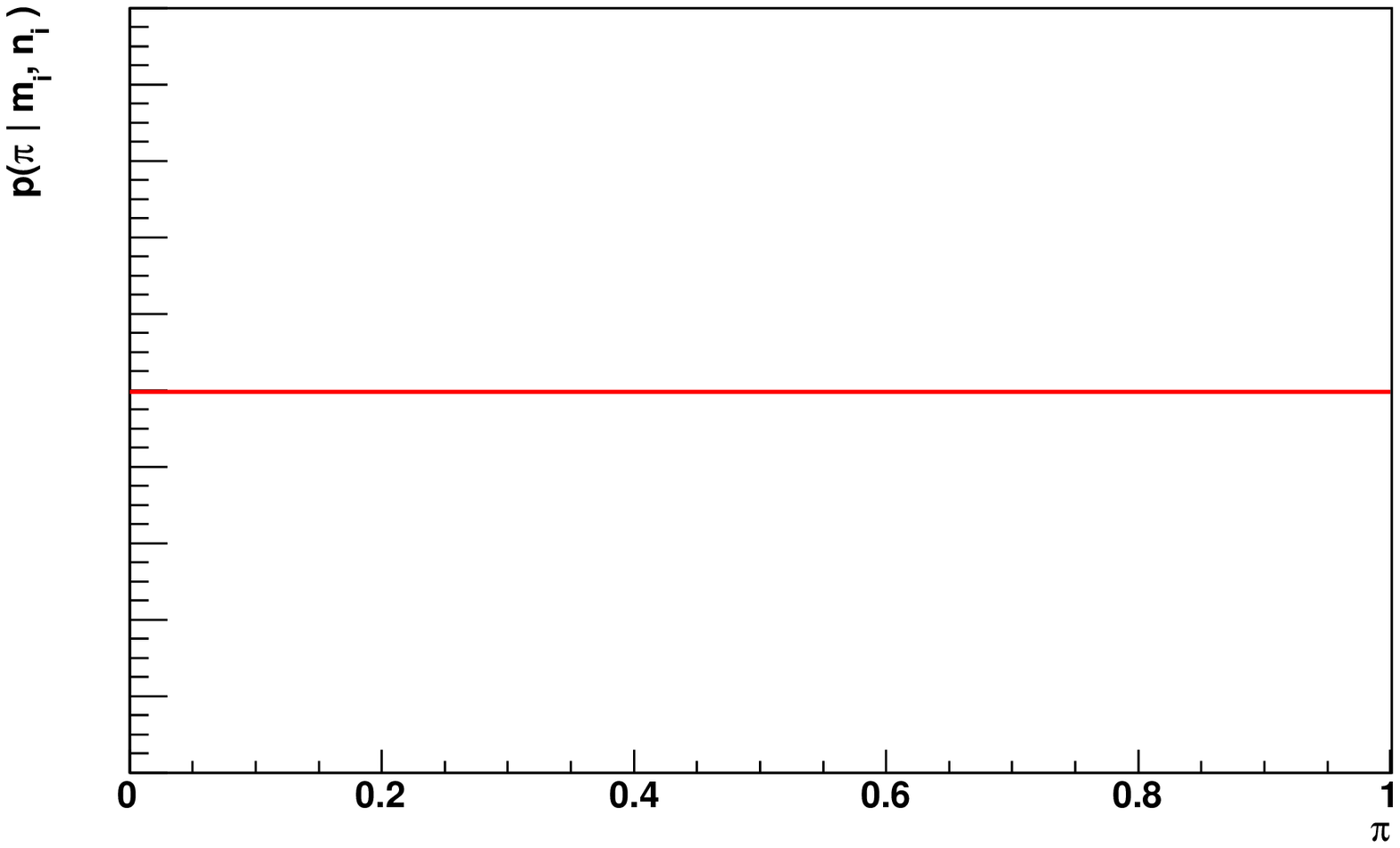}}
\subfigure[]{\includegraphics[width=1.9in]{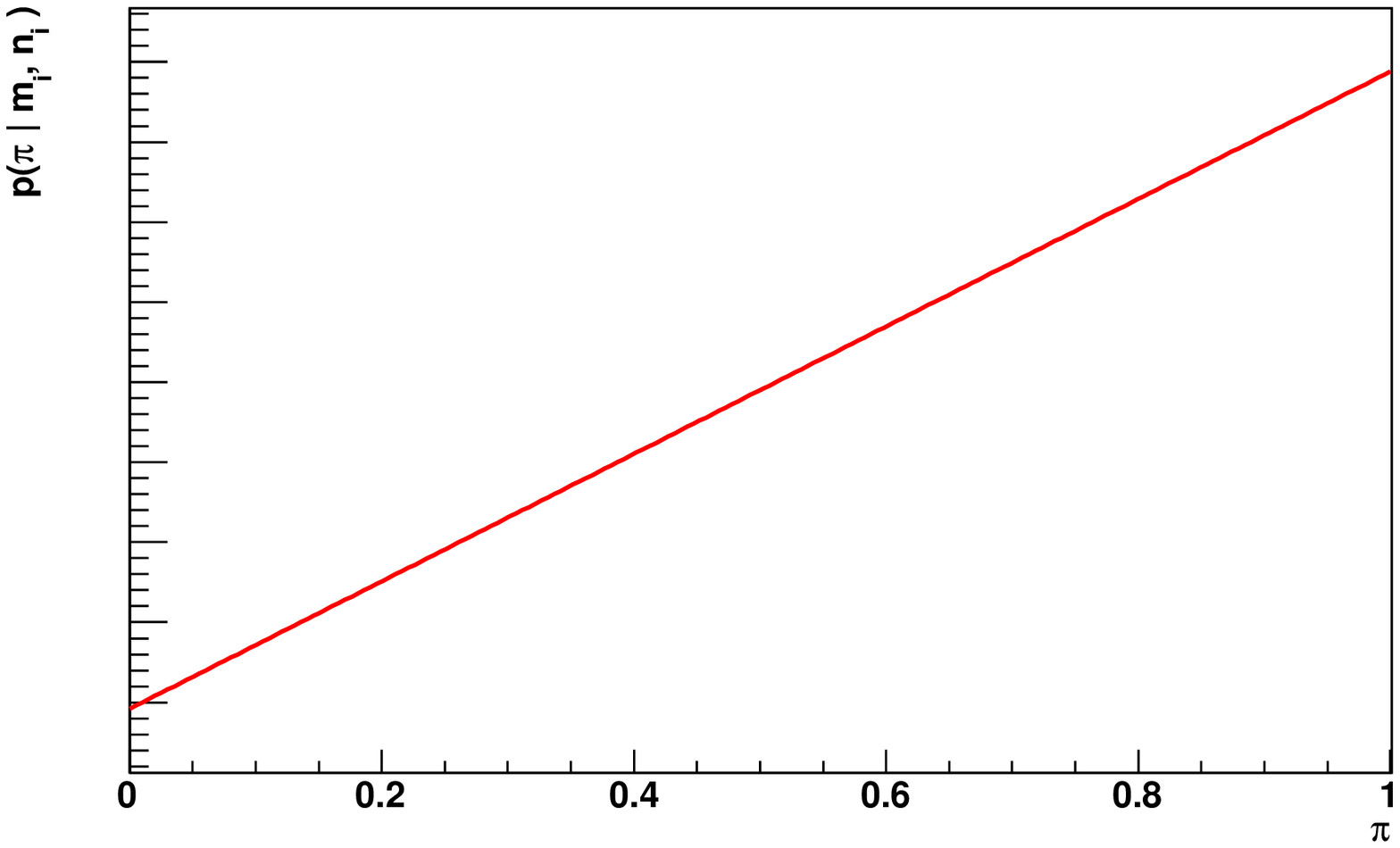}}
\caption{ The three classes of mixture posterior for a single bin: (a) favoring $\bar{\mathcal{S}}$, (b) indifferent between the two models, and (c) favoring $\mathcal{S}$ .
\label{fig:single}}
\end{figure}

In the case of a single bin and a uniform prior, the mixture posterior is linear in $\pi$ and the possible posteriors fall into three classes.  Those with a negative slope are best modeled by $\bar{\mathcal{S}}$ (Fig \ref{fig:single}a), those uniform across $\pi$ are indifferent between the two models (Fig \ref{fig:single}b), and those with a positive slope favor $\mathcal{S}$ (Fig \ref{fig:single}c).

\begin{figure}
\centering
\subfigure[]{\includegraphics[width=1.9in]{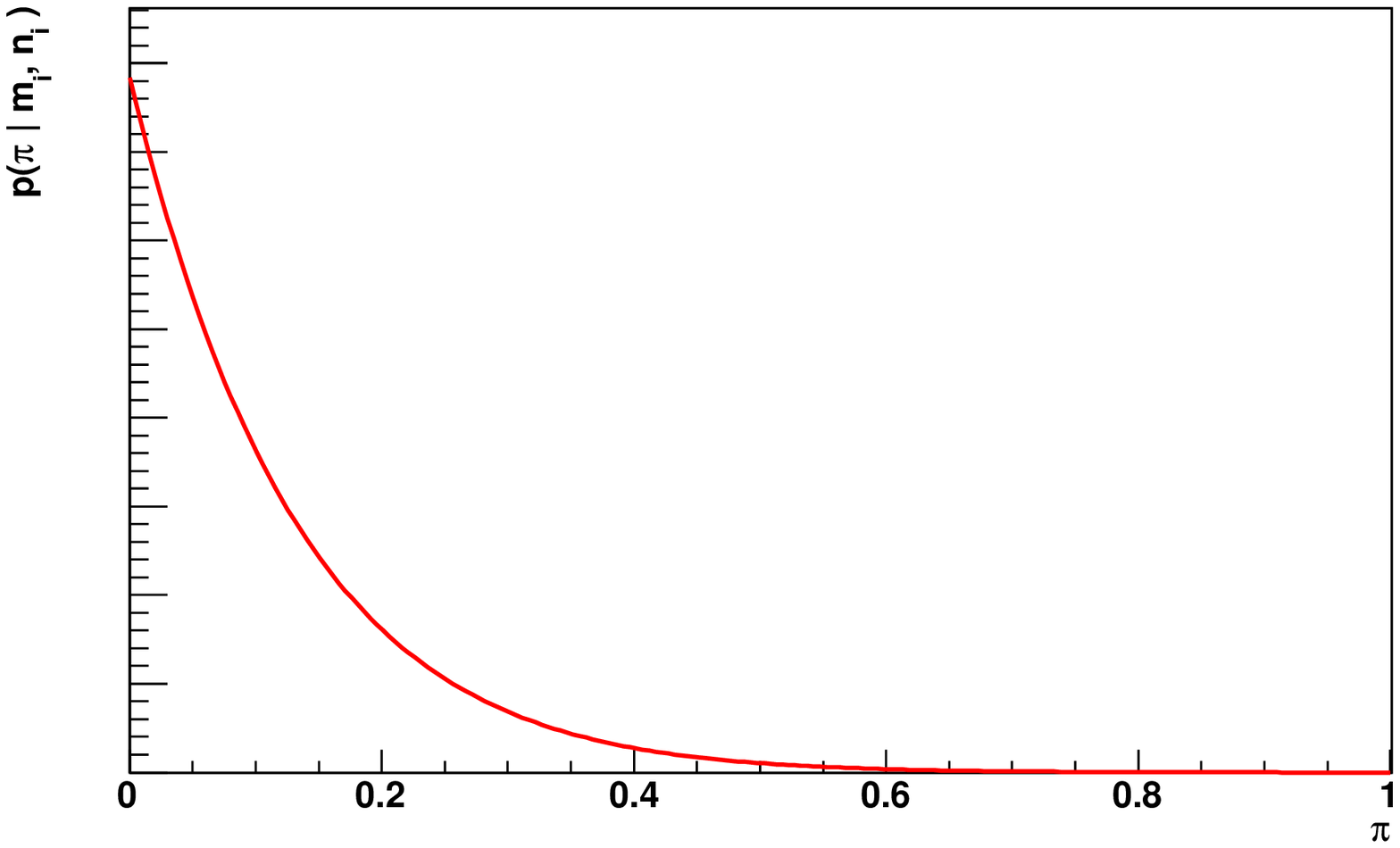}}
\subfigure[]{\includegraphics[width=1.9in]{singleFlat.eps}}
\subfigure[]{\includegraphics[width=1.9in]{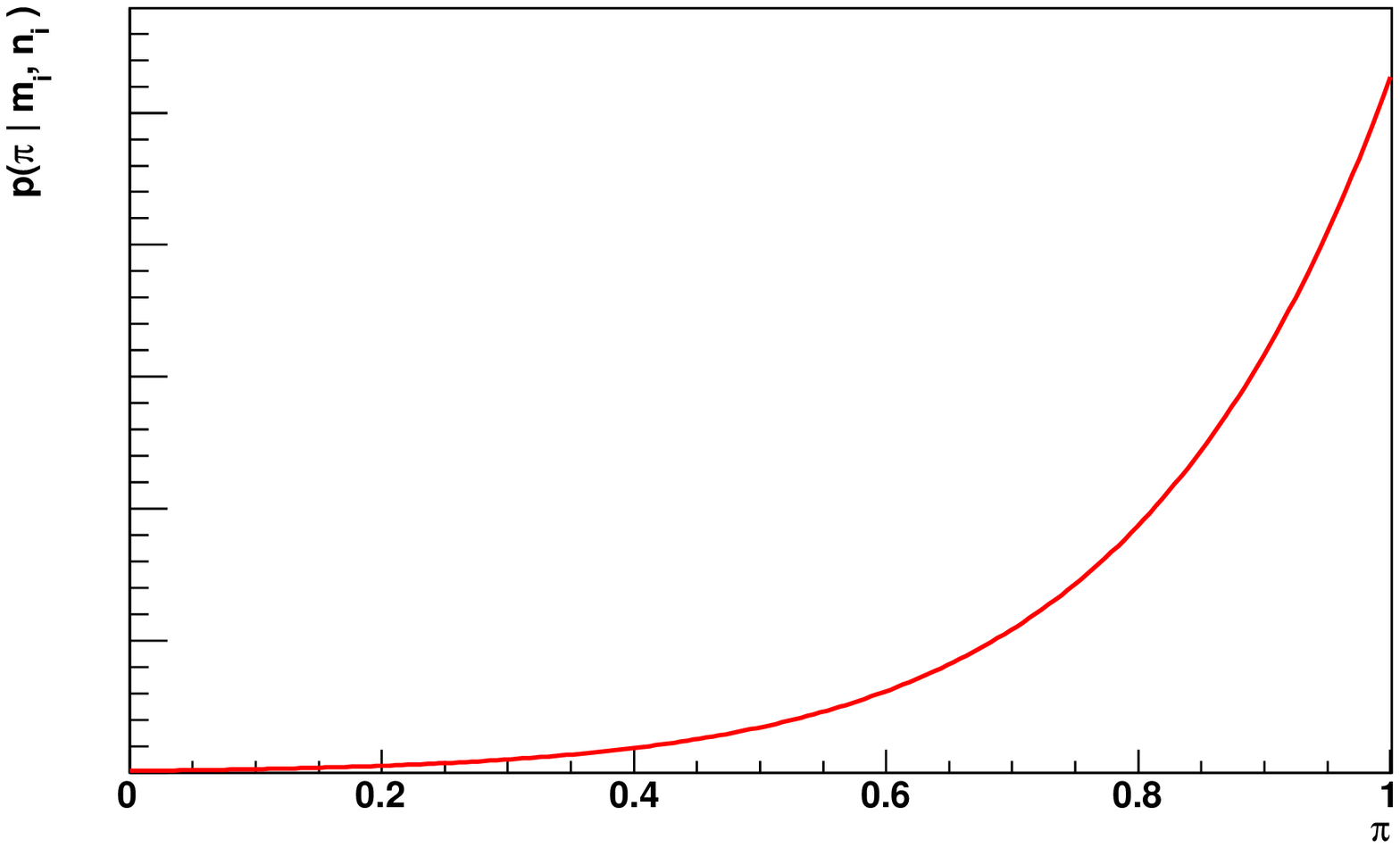}}
\caption{ When all bins agree the posterior is either (a) concentrated towards $\pi = 0$ and favoring $\bar{\mathcal{S}}$ , (b) indifferent, or (c) concentrated towards $\pi = 1$ and favoring $\mathcal{S}$ .
\label{fig:mult}}
\end{figure}

\begin{figure}
\centering
\includegraphics[width=3in]{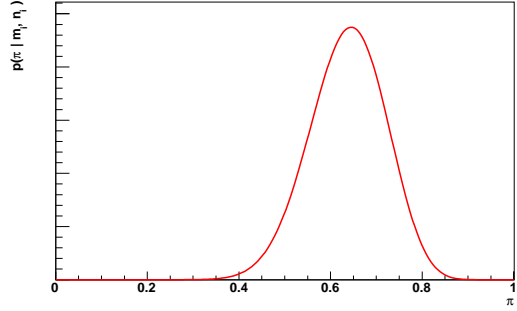}
\caption{ A local mode arises when individual bins are best described by different models.
\label{fig:multMixed}}
\end{figure}

In the general case, each bin is modeled independently.  If all bins favor $\bar{\mathcal{S}}$ then the combined product quickly concentrates towards small $\pi$ with a mode exactly at $\pi = 0$ (Fig \ref{fig:mult}a).  Conversely, full support of $\mathcal{S}$ produces a posterior concentrated towards $\pi = 1$ with a mode exactly at the boundary (Fig \ref{fig:mult}c).  Any bins indifferent between the models contribute little to the final shape, unless all the bins are ignorant in which case the posterior remains uniform (Fig \ref{fig:mult}b).  When bins are best modeled by different models, however, the posterior becomes peaked away from the boundaries with the exact location of the mode indicating the overall superior model (Fig \ref{fig:multMixed}).

Diagnosing the best model is then straightforward.  When a mode arises at the boundaries, the bins are consistently described by a single model: $\bar{\mathcal{S}}$ for $\pi = 0$ and $\mathcal{S}$ for $\pi = 1$.  A mode away from the boundaries indicates mixed agreement, implying that the two histograms are not entirely consistent.  

\pagebreak

Note that the non-mixture posterior can be recovered by comparing the boundaries of the mixture posterior,

\begin{equation*}
p \left( \mathcal{S} | \mathbf{m}, \mathbf{n} \right) = \frac{ p \left( \pi = 1 | \mathbf{m}, \mathbf{n} \right) }{ p \left( \pi = 0 | \mathbf{m}, \mathbf{n} \right) + p \left( \pi = 1 | \mathbf{m}, \mathbf{n} \right)}
\end{equation*}

The additional information contained in the shape of mixture posterior underlies the benefit of the extension to a mixture model.  Not only does the mixture model break the the degeneracy between ignorance and balancing disagreement, but it can also reveal more subtle disagreements that might otherwise be obscured in the model posterior.  

\subsection{Comparing Data}

Actually computing the mixture posterior requires the bin by bin evidences, $p \left( m_{i}, n_{i} | \mathcal{S} \right)$ and $p \left( m_{i}, n_{i} | \bar{\mathcal{S}} \right)$.

\subsubsection{The Same Source Model Evidence}

When both bin populations are drawn from the same distribution, the joint likelihood is given by

\begin{align*}
p \left( m_{i}, n_{i} | \lambda, \mathcal{S} \right) &= p \left( m_{i} | \lambda  \right) p \left( n_{i} | \lambda  \right) \\
 &= \frac{ \lambda^{m_{i}} e^{- \lambda} }{m_{i}!} \frac{ \lambda^{n_{i}} e^{- \lambda} }{n_{i}!} \\
 &= \frac{ \lambda^{m_{i} + n_{i}} e^{- 2 \lambda} }{m_{i}! n_{i}!}. \\
\end{align*}

Assigning a uniform prior,

\begin{equation*}
p \left( \lambda | \mathcal{S} \right) = \left\{ \begin{array}{rc} \left( \beta - \alpha \right)^{-1} ,& \alpha \leq \lambda \leq \beta \\ 0 ,& \mathrm{else} \\ \end{array} \right. ,
\end{equation*}

\noindent yields the posterior

\begin{align*}
p \left( \lambda | m_{i}, n_{i}, \mathcal{S} \right) &\propto p \left( m_{i}, n_{i} | \lambda, \mathcal{S} \right) p \left( \lambda | \mathcal{S} \right) \\
&= \frac{ \lambda^{m_{i} + n_{i}} e^{-2 \lambda} }{ \int_{\alpha}^{\beta} \mathrm{d} \lambda \, \lambda^{m_{i} + n_{i}} e^{-2 \lambda} } \\
&= \frac{ \lambda^{m_{i} + n_{i}} e^{-2 \lambda} }{ 2^{- \left( m_{i} + n_{i} + 1 \right) } \Gamma \left( m_{i} + n_{i} + 1 \right) \left[ \bar{\gamma} \left( m_{i} + n_{i} + 1, 2 \beta \right) - \bar{\gamma} \left( m_{i} + n_{i} + 1, 2 \alpha \right) \right] },
\end{align*}

\noindent where $\bar{\gamma}$ is the normalized lower incomplete gamma function,

\begin{equation*}
\bar{\gamma} \left(n, x \right) = \frac{ \int_{0}^{x} \mathrm{d}t \, t^{n - 1} e^{-t} }{ \int_{0}^{\infty} \mathrm{d}t \, t^{n - 1} e^{-t} } .
\end{equation*}

The evidence follows,

\begin{align*}
p \left( m_{i}, n_{i} | \mathcal{S} \right) &= \frac{ p \left( m_{i}, n_{i} | \lambda, \mathcal{S} \right) p \left( \lambda, \mathcal{S} \right) }{ p \left( \lambda | m_{i}, n_{i}, \mathcal{S} \right) }\\
 &= \frac{1}{ \beta - \alpha } \frac{2^{- \left(m_{i}  + n_{i} + 1 \right)}}{m_{i} + n_{i} + 1 } \frac{\bar{\gamma} \left( m_{i} + n_{i} + 1, 2 \beta \right) - \bar{\gamma} \left( m_{i} + n_{i} + 1, 2 \alpha \right)}{\mathrm{Be} \left( m_{i} + 1, n_{i} + 1 \right) }, \\
\end{align*}

\noindent where $\mathrm{Be} \left( x, y \right)$ is the Beta function,

\begin{equation*}
\mathrm{Be} \left( x, y \right) = \int_{0}^{1} \mathrm{d} t \, t^{x} \left(1 - t\right)^{1 - y}.
\end{equation*}

In practice the prior support, $\alpha \leq \lambda \leq \beta$, is chosen to encompass the bulk of the likelihood.  Making the support as small as possible while keeping the posterior normalization practically constant implements a crude form of model selection that improves the model without limiting its power.

\pagebreak

\subsubsection{The Two Source Model Evidence}

Given two distinct sources, the joint likelihood factors into

\begin{align*}
p \left( m_{i}, n_{i} | \lambda, \bar{ \mathcal{S} } \right) &= p \left( m_{i} | \lambda  \right) \cdot p \left( n_{i} | \mu  \right) \\
 &= \frac{ \lambda^{m_{i}} e^{- \lambda} }{m_{i}!} \cdot \frac{ \mu^{n_{i}} e^{- \mu} }{n_{i}!} .
\end{align*}

The two sources require two independent priors,

\begin{align*}
p \left( \lambda, \mu | \bar{ \mathcal{S} } \right) &= p \left( \lambda | \bar{ \mathcal{S} } \right) \cdot p \left( \mu | \bar{ \mathcal{S} } \right) \\
 &= \left\{ \begin{array}{rc}  \left( \delta - \gamma  \right)^{-1} \cdot \left( \zeta - \epsilon \right)^{-1} ,& \gamma \leq \lambda \leq \delta , \epsilon \leq \mu \leq \zeta \\ 0 ,& \mathrm{else} \\ \end{array} \right. ,
\end{align*}

\noindent and the joint posterior becomes

\begin{align*}
p \left( \lambda, \mu | m_{i}, n_{i}, \bar{ \mathcal{S} } \right) &= p \left( \lambda | m_{i}, n_{i}, \bar{ \mathcal{S} } \right) \cdot p \left( \mu | m_{i}, n_{i}, \bar{ \mathcal{S} } \right) \\
 &\propto p \left( m_{i}, n_{i} | \lambda, \bar{ \mathcal{S} } \right) p \left( \lambda | \bar{ \mathcal{S} } \right) \cdot p \left( m_{i}, n_{i} | \mu, \bar{ \mathcal{S} } \right) p \left( \mu | \bar{ \mathcal{S} } \right) \\
 &= \frac{ \lambda^{m_{i}} e^{-\lambda} }{ \int_{\gamma}^{\delta} \mathrm{d} \lambda\, \lambda^{m_{i}} e^{-\lambda} } \frac{ \mu^{n_{i}} e^{- \mu} }{ \int_{\epsilon}^{\zeta} \mathrm{d} \mu \, \mu^{n_{i}} e^{- \mu} } \\
 &= \frac{ \lambda^{m_{i}} e^{-\lambda} }{ \Gamma \left(m_{i} + 1 \right) \left[ \bar{ \gamma } \left( m_{i} + 1, \delta \right) - \bar{ \gamma } \left( m_{i} + 1, \gamma \right) \right] } \frac{ \mu^{n_{i}} e^{- \mu} }{ \Gamma \left(n_{i} + 1 \right) \left[ \bar{ \gamma } \left( n_{i} + 1, \zeta \right) - \bar{ \gamma } \left( n_{i} + 1, \epsilon \right) \right] } .
\end{align*}

\noindent The evidence is then 

\begin{align*}
p \left( m_{i}, n_{i} | \bar{ \mathcal{S} } \right) &= \frac{ p \left( m_{i}, n_{i} | \lambda, \mu \bar{ \mathcal{S} } \right) p \left( \lambda, \mu \bar{ \mathcal{S} } \right) }{ p \left( \lambda, \mu | m_{i}, n_{i}, \bar{ \mathcal{S} } \right) }\\
 &= \frac{ p \left( m_{i}, n_{i} | \lambda, \bar{ \mathcal{S} } \right) p \left( \lambda \bar{ \mathcal{S} } \right) }{ p \left( \lambda | m_{i}, n_{i}, \bar{ \mathcal{S} } \right) } \cdot \frac{ p \left( m_{i}, n_{i} | \mu, \bar{ \mathcal{S} } \right) p \left( \mu \bar{ \mathcal{S} } \right) }{ p \left( \mu | m_{i}, n_{i}, \bar{ \mathcal{S} } \right) } \\
 &=  \frac{ \Gamma \left(m_{i} + 1 \right) \left[ \bar{ \gamma } \left( m_{i} + 1, \delta \right) - \bar{ \gamma } \left( m_{i} + 1, \gamma \right) \right] }{  \Gamma \left( m_{i} + 1 \right) \left( \delta - \gamma \right)  } \frac{ \Gamma \left(n_{i} + 1 \right) \left[ \bar{ \gamma } \left( n_{i} + 1, \zeta \right) - \bar{ \gamma } \left( n_{i} + 1, \epsilon \right) \right] }{  \Gamma \left( n_{i} + 1\right) \left( \zeta - \epsilon \right)  } \\
 &= \frac{ \left[ \bar{ \gamma } \left( m_{i} + 1, \delta \right) - \bar{ \gamma } \left( m_{i} + 1, \gamma \right) \right] }{ \left( \delta - \gamma \right)  } \frac{ \left[ \bar{ \gamma } \left( n_{i} + 1, \zeta \right) - \bar{ \gamma } \left( n_{i} + 1, \epsilon \right) \right] }{ \left( \zeta - \epsilon \right)  } .
\end{align*}

\subsection{Importance Sampling}

\subsubsection{Sampling as Monte Carlo Estimation}

Cross sections defining the physics from which the data are generated are equivalent to the probability distribution

\begin{equation*}
p \left( x \right) = \frac{1}{\sigma } \frac{ d \sigma \left( x \right) }{dx},
\end{equation*}

\noindent where $x$ defines the phase space of the given physics.

Given an integrated luminosity $\bar{\mathcal{L}}$, the expected number of events falling into a particular histogram bin is then

\begin{align*}
\lambda &= \int_{a}^{b} \mathrm{d}x \, \bar{\mathcal{L}}  \frac{ d \sigma \left( x \right) }{dx} \\
&= \int_{a}^{b} \mathrm{d}x \, \bar{\mathcal{L}} \sigma p \left( x\right) \\
&= \int_{-\infty}^{\infty} \mathrm{d}x \, I_{ab} \left( x\right) \bar{\mathcal{L}} \sigma \, p \left( x\right) \\
&\equiv \int_{-\infty}^{\infty} \mathrm{d}x N \left(x\right) p \left( x\right),
\end{align*}

\noindent where

\begin{equation*}
I_{ab} \left( x\right) = \left\{ \begin{array}{rc} 1 ,&  a \leq x \leq b \\ 0 ,& \mathrm{else} \end{array} \right. .
\end{equation*}

Now consider a Monte Carlo simulation generating $n = \bar{\mathcal{L}} \sigma$ events from the differential cross section $d \sigma / dx$.  The samples admit the Monte Carlo estimate of the expected counts, 

\begin{align*}
\tilde{N} &= \frac{1}{n} \sum_{i=1}^{n} N \left(x_{i} \right) \\
&= \frac{1}{n} \sum_{i = 1}^{n} \bar{\mathcal{L}} \sigma I_{ab} \left( x_{i} \right) \\
&= \sum_{i = 1}^{n} I_{ab} \left( x_{i} \right), 
\end{align*}

\noindent which is just the number of events falling into the bin $a \leq x \leq b$.

\pagebreak

The estimate is unbiased, 

\begin{equation*}
\mathrm{E}_{p} [ \tilde{N} ] = \lambda,
\end{equation*}

\noindent with the variance

\begin{align*}
\mathrm{Var}_{p} [ \tilde{N} ] &= \frac{1}{n} \mathrm{Var}_{p} \left[ N \left( x \right) \right] \\
&= \frac{1}{n} \left( \mathrm{E}_{p} \left[ N^{2} \left(x\right) \right] - \mathrm{E}_{p} \left[ N \left(x\right) \right]^{2} \right) \\
&= \frac{1}{n} \left( \int_{-\infty}^{\infty} I^{2}_{ab} \left(x\right) \left( \bar{\mathcal{L}} \sigma\right)^{2} \, p \left(x\right) - \lambda^{2} \right) \\
&= \frac{1}{n} \left( \bar{\mathcal{L}} \sigma \int_{-\infty}^{\infty} I_{ab} \left(x\right) \bar{\mathcal{L}} \sigma \, p \left(x\right) - \lambda^{2} \right) \\
&= \frac{1}{n} \left( n \lambda - \lambda^{2} \right) \\
&= \lambda - \lambda^{2} / n .
\end{align*}

In the large $n$ limit, the second term becomes negligible and the sampling distribution of the Monte Carlo estimate converges to a Poisson distribution,

\begin{equation*}
p \left( \tilde{N} | \lambda \right) = \frac{ \lambda^{\tilde{N}} e^{-\lambda} }{ \tilde{N}! } .
\end{equation*}

\noindent Note that this limit does not require large bin contents.  In fact, the large $n$ limit is exactly the same rare-events limit in which the multinomial distribution reduces to independent Poisson distributions.  The Monte Carlo samples, then, are entirely analogous to the data and inferring the underlying rate $\lambda$ is the same in both cases.

Importance sampling is slightly more subtle.  Here events are not drawn from the physics distribution $p \left( x\right)$ but instead an auxiliary distribution $q \left( x\right)$.  The expected number of events becomes

\begin{align*}
\lambda &= \int_{-\infty}^{\infty} \mathrm{d}x \, N \left(x\right) p \left( x\right) \\
&= \int_{-\infty}^{\infty} \mathrm{d}x N \left(x\right) \frac{p \left( x\right) }{ q \left( x \right)} q \left( x \right) \\
&\equiv \int_{-\infty}^{\infty} \mathrm{d}x \, N \left(x\right) w \left( x\right) q \left( x \right) \\
\end{align*}

\pagebreak

\noindent with the resulting Monte Carlo estimate

\begin{align*}
\tilde{N} &= \frac{1}{n} \sum_{i=1}^{n} N \left(x_{i}\right) w \left(x_{i}\right) \\ 
&= \frac{1}{n} \sum_{i = 1}^{n} \bar{\mathcal{L}} \sigma I_{ab} \left( x_{i} \right) w \left( x_{i} \right) \\
&= \frac{\bar{\mathcal{L}} \sigma}{n} \sum_{i = 1}^{n} I_{ab} \left( x_{i} \right) w \left( x_{i} \right) \\
&= \sum_{i = 1}^{n} I_{ab} \left( x_{i} \right)  w \left( x_{i} \right).
\end{align*}

\noindent Note that this estimate is just the sum of the weights of events falling into the chosen bin.

As above, the importance sampling estimate is an unbiased estimator,

\begin{equation*}
\mathrm{E}_{q} [ \tilde{N} ] = \lambda,
\end{equation*}

\noindent but with the variance

\begin{align*}
\mathrm{Var}_{q} [ \tilde{N} ] &= \frac{1}{n} \mathrm{Var}_{q} \left[ N \left(x\right) w \left(x\right) \right] \\
&= \frac{1}{n} \left( \mathrm{E}_{q} \left[ N^{2} \left(x\right) w^{2} \left(x\right) \right] - \mathrm{E}_{q} \left[ N \left(x\right) w \left(x\right) \right]^{2} \right) \\
&= \frac{1}{n} \left( \int_{-\infty}^{\infty} I^{2}_{ab} \left(x\right) \left( \bar{\mathcal{L}} \sigma\right)^{2} w^{2} \left( x\right) \, q \left(x\right) - \lambda^{2} \right) \\
&= \frac{1}{n} \left( \bar{\mathcal{L}} \sigma \int_{-\infty}^{\infty} I_{ab} \left(x\right) \bar{\mathcal{L}} \sigma \,  w \left( x\right) p \left(x\right) - \lambda^{2} \right) \\
&= \frac{1}{n} \left( n \int_{-\infty}^{\infty} I_{ab} \left(x\right) \bar{\mathcal{L}} \sigma \,  w \left( x\right) p \left(x\right) - \lambda^{2} \right) \\
&= \int_{-\infty}^{\infty} I_{ab} \left(x\right) \bar{\mathcal{L}} \sigma \,  w \left( x\right) p \left(x\right) - \lambda^{2} / n .
\end{align*}

If the weighting function $w \left(x\right)$ is relatively constant across the bin $a \leq x \leq b$ then $w \left(x\right)$ can be approximated by its ensemble average,

\begin{equation*}
\bar{w} = \frac{ \sum_{i=1}^{n} I_{ab} \left( x_{i} \right) w \left( x_{i} \right) }{ \sum_{i=1}^{n} I_{ab} \left( x_{i} \right) },
\end{equation*}

\pagebreak

\noindent and the remaining integral becomes

\begin{align*}
\int_{-\infty}^{\infty} I_{ab} \left(x\right) \bar{\mathcal{L}} \sigma \,  w \left( x\right) p \left(x\right) &\approx \int_{-\infty}^{\infty} I_{ab} \left(x\right) \bar{\mathcal{L}} \sigma \,  \bar{w} p \left(x\right) \\
&\approx \bar{w} \int_{-\infty}^{\infty} I_{ab} \left(x\right) \bar{\mathcal{L}} \sigma \,  p \left(x\right) \\
&\approx \bar{w} \lambda .
\end{align*}

The variance is then

\begin{align*}
\mathrm{Var}_{q} [ \tilde{N} ] &\approx \bar{w} \lambda - \lambda^{2} / n, 
\end{align*}

\noindent and in the large $n$ limit reduces to

\begin{equation*}
\mathrm{Var}_{q} [ \tilde{N} ] \approx \bar{w} \lambda .
\end{equation*}

\noindent Note that if the weights are unity, in which case the importance sampling reduces to standard Monte Carlo sampling, then the variance reduces to the previous result as consistency would demand.

The resulting sampling distribution is approximately Gaussian,

\begin{equation*}
p \left( \tilde{N} | \lambda \right) = \frac{1}{\sqrt{2\pi \bar{w} \lambda}} \exp\left[ - \frac{1}{2} \frac{ \left(\tilde{N} - \lambda\right)^{2} }{\bar{w} \lambda } \right].
\end{equation*}

The above sampling distribution admits a Bayesian approach to simulated histograms and consequently the comparison of simulated histograms.  Computation of the evidence ratio follows as in the case of data comparison.

\subsection{Comparing Data to Simulation}

The importance sampling distribution for the simulated histogram bins requires not only the sum of the weights falling into that bin,

\begin{equation*}
n_{i} = \sum_{j = 1}^{n} I_{a_{i} b_{i} } \left( x_{j} \right)  w \left( x_{j} \right),
\end{equation*}

\noindent but also the ensemble average of the weights, 

\begin{equation*}
\bar{w}_{i} = \frac{ \sum_{j=1}^{n} I_{a_{i} b_{i} } \left( x_{j} \right) w \left( x_{j} \right) }{ \sum_{j=1}^{n} I_{a_{i} b_{i} } \left( x_{j} \right) }.
\end{equation*}

As before, the sampling distribution of the data histograms bins $m_{i}$ is taken to be Poisson.

\subsubsection{The Same Source Model Evidence}

Here the joint likelihood becomes

\begin{align*}
p \left( m_{i}, n_{i} | \lambda, \mathcal{S} \right) &= p \left( m_{i} | \lambda \right) p \left( n_{i} | \lambda \right) \\
&= \frac{\lambda^{m_{i}} e^{-\lambda}}{m_{i}!} \frac{1}{\sqrt{ 2 \pi \bar{w}_{i} \lambda } } \exp \left[ - \frac{1}{2} \frac{ \left( n_{i} - \lambda \right)^{2} }{\bar{w}_{i} \lambda} \right] .
\end{align*}

A uniform prior,

\begin{equation*}
p \left( \lambda | \mathcal{S} \right)  = \left\{ \begin{array}{rc}  \left( \beta - \alpha \right)^{-1},& \alpha \leq \lambda \leq \beta \\ 0,& \mathrm{else} \end{array} \right. ,
\end{equation*}

\noindent gives the posterior

\begin{align*}
p \left( \lambda | m_{i}, n_{i}, \mathcal{S} \right) &\propto p \left( m_{i}, n_{i} | \lambda, \mathcal{S} \right) p \left( \lambda |  \mathcal{S} \right) \\
&= \frac{ \lambda^{m_{i} - \frac{1}{2} } \exp \left[ \lambda - \frac{1}{2} \frac{ \left( n_{i} - \lambda \right)^{2} }{\bar{w}_{i} \lambda} \right] }{ \int_{\alpha}^{\beta} \mathrm{d} \lambda \, \lambda^{m_{i} - \frac{1}{2} } \exp \left[ \lambda - \frac{1}{2} \frac{ \left( n_{i} - \lambda \right)^{2} }{\bar{w}_{i} \lambda} \right] },
\end{align*}

\noindent resulting in the evidence

\begin{equation*}
p \left( m_{i}, n_{i} | \mathcal{S} \right) = \frac{1}{ m_{i}! \left( \beta - \alpha \right) \sqrt{ 2\pi \bar{w}_{i} } } \int_{\alpha}^{\beta} \mathrm{d} \lambda \, \lambda^{m_{i} - \frac{1}{2} } \exp \left[ \lambda - \frac{1}{2} \frac{ \left( n_{i} - \lambda \right)^{2} }{\bar{w}_{i} \lambda} \right].
\end{equation*}

Note that limiting the prior support not only improves the model but also eases the computational demands of the required numerical integration.

\subsubsection{The Two Source Model Evidence}

Given two distinct sources, the joint likelihood factors into

\begin{align*}
p \left( m_{i}, n_{i} | \lambda, \bar{ \mathcal{S} } \right) &= p \left( m_{i} | \lambda  \right) \cdot p \left( n_{i} | \mu  \right) \\
 &= \frac{ \lambda^{m_{i}} e^{- \lambda} }{m_{i}!} \cdot \frac{1}{\sqrt{ 2 \pi \bar{w}_{i} \lambda } } \exp \left[ - \frac{1}{2} \frac{ \left( n_{i} - \lambda \right)^{2} }{\bar{w}_{i} \lambda} \right] .
\end{align*}

\pagebreak

The two sources require two independent priors,

\begin{align*}
p \left( \lambda, \mu \bar{ \mathcal{S} } \right) &= p \left( \lambda | \bar{ \mathcal{S} } \right) \cdot p \left( \mu | \bar{ \mathcal{S} } \right) \\
 &= \left\{ \begin{array}{rc}  \left( \delta - \gamma  \right)^{-1} \cdot \left( \zeta - \epsilon \right)^{-1} ,& \gamma \leq \lambda \leq \delta , \epsilon \leq \mu \leq \zeta \\ 0 ,& \mathrm{else} \\ \end{array} \right. ,
\end{align*}

\noindent and the joint posterior becomes

\begin{align*}
p \left( \lambda, \mu | m_{i}, n_{i}, \bar{ \mathcal{S} } \right) &= p \left( \lambda | m_{i}, n_{i}, \bar{ \mathcal{S} } \right) \cdot p \left( \mu | m_{i}, n_{i}, \bar{ \mathcal{S} } \right) \\
 &\propto p \left( m_{i}, n_{i} | \lambda, \bar{ \mathcal{S} } \right) p \left( \lambda | \bar{ \mathcal{S} } \right) \cdot p \left( m_{i}, n_{i} | \mu, \bar{ \mathcal{S} } \right) p \left( \mu | \bar{ \mathcal{S} } \right) \\
 &= \frac{ \lambda^{m_{i}} e^{- \lambda} }{\int_{\gamma}^{\delta} \mathrm{d} \lambda \, \lambda^{m_{i}} e^{- \lambda}} \frac{\mu^{-\frac{1}{2}} \exp \left[ - \frac{1}{2} \frac{ \left( n_{i} - \mu \right)^{2} }{\bar{w}_{i} \mu} \right] }{ \int_{\epsilon}^{\zeta}  \mathrm{d} \mu \, \mu^{-\frac{1}{2}} \exp \left[ - \frac{1}{2} \frac{ \left( n_{i} - \mu \right)^{2} }{\bar{w}_{i} \mu} \right] } \\
 &= \frac{ \lambda^{m_{i}} e^{- \lambda} }{  \Gamma \left(m_{i} + 1 \right) \left[ \bar{\gamma} \left( m_{i} + 1, \delta \right) - \bar{\gamma} \left( m_{i} + 1, \gamma \right) \right] } \frac{ \mu^{-\frac{1}{2}} \exp \left[ - \frac{1}{2} \frac{ \left( n_{i} - \mu \right)^{2} }{\bar{w}_{i} \mu} \right] }{ \int_{\epsilon}^{\zeta} \mathrm{d} \mu \, \mu^{-\frac{1}{2}} \exp \left[ - \frac{1}{2} \frac{ \left( n_{i} - \mu \right)^{2} }{\bar{w}_{i} \mu} \right] } ,
\end{align*}

\noindent yielding the evidence

\begin{align*}
p \left( m_{i}, n_{i} | \bar{ \mathcal{S} } \right) &= \frac{ p \left( m_{i}, n_{i} | \lambda, \mu,  \bar{ \mathcal{S} } \right) p \left( \lambda, \mu, \bar{ \mathcal{S} } \right) }{ p \left( \lambda, \mu | m_{i}, n_{i}, \bar{ \mathcal{S} } \right) }\\
 &= \frac{ p \left( m_{i}, n_{i} | \lambda, \bar{ \mathcal{S} } \right) p \left( \lambda | \bar{ \mathcal{S} } \right) }{ p \left( \lambda | m_{i}, n_{i}, \bar{ \mathcal{S} } \right) } \cdot \frac{ p \left( m_{i}, n_{i} | \mu, \bar{ \mathcal{S} } \right) p \left( \mu | \bar{ \mathcal{S} } \right) }{ p \left( \mu | m_{i}, n_{i}, \bar{ \mathcal{S} } \right) } \\
&= \frac{\left( \delta - \gamma \right)^{-1} \left( \zeta - \epsilon \right)^{-1}}{m_{i}! \sqrt{2\pi \bar{w}_{i} } } \Gamma \left(m_{i} + 1 \right) \left[ \bar{\gamma} \left( m_{i} + 1, \delta \right) - \bar{\gamma} \left( m_{i} + 1, \gamma \right) \right]\cdot \int_{\epsilon}^{\zeta} \mathrm{d} \mu \, \mu^{-\frac{1}{2}} \exp \left[ - \frac{1}{2} \frac{ \left( n_{i} - \mu \right)^{2} }{\bar{w}_{i} \mu} \right] 
\end{align*}

\subsection{Comparing Simulation to Simulation}

Comparing two simulated histogram bins requires the sum of the weights and ensemble average for both the first bin,

\begin{equation*}
m_{i} = \sum_{j = 1}^{n} I_{a_{i} b_{i} } \left( x_{j} \right)  w \left( x_{j} \right)
\end{equation*}

\begin{equation*}
\bar{w}_{i} = \frac{ \sum_{j=1}^{n} I_{a_{i} b_{i} } \left( x_{j} \right) w \left( x_{j} \right) }{ \sum_{j=1}^{n} I_{a_{i} b_{i} } \left( x_{j} \right) } ,
\end{equation*}

\pagebreak

\noindent and the second bin,

\begin{equation*}
n_{i} = \sum_{j = 1}^{n} I_{a_{i} b_{i} } \left( x_{j} \right)  v \left( x_{j} \right)
\end{equation*}

\begin{equation*}
\bar{v}_{i} = \frac{ \sum_{j=1}^{n} I_{a_{i} b_{i} } \left( x_{j} \right) v \left( x_{j} \right) }{ \sum_{j=1}^{n} I_{a_{i} b_{i} } \left( x_{j} \right) }.
\end{equation*}

\subsubsection{The Same Source Model Evidence}

In this case the joint likelihood becomes

\begin{align*}
p \left( m_{i}, n_{i} | \lambda, \mathcal{S} \right) &= p \left( m_{i} | \lambda \right) p \left( n_{i} | \lambda \right) \\
&= \frac{1}{\sqrt{ 2 \pi \bar{w}_{i} \lambda } } \exp \left[ - \frac{1}{2} \frac{ \left( m_{i} - \lambda \right)^{2} }{\bar{w}_{i} \lambda} \right] \frac{1}{\sqrt{ 2 \pi \bar{v}_{i} \lambda } } \exp \left[ - \frac{1}{2} \frac{ \left( n_{i} - \lambda \right)^{2} }{\bar{v}_{i} \lambda} \right] \\
&= \frac{1}{ 2 \pi \lambda \sqrt{ \bar{w}_{i} \bar{v}_{i} } } \exp \left[ - \frac{1}{2 \lambda} \left( \frac{ \left( m_{i} - \lambda \right)^{2} }{\bar{w}_{i} } - \frac{ \left( n_{i} - \lambda \right)^{2} }{\bar{v}}_{i} \right) \right]
\end{align*}

A uniform prior,

\begin{equation*}
p \left( \lambda | \mathcal{S} \right)  = \left\{ \begin{array}{rc}  \left( \beta - \alpha \right)^{-1},& \alpha \leq \lambda \leq \beta \\ 0,& \mathrm{else} \end{array} \right. ,
\end{equation*}

\noindent gives the posterior

\begin{align*}
p \left( \lambda | m_{i}, n_{i}, \mathcal{S} \right) &\propto p \left( m_{i}, n_{i} | \lambda, \mathcal{S} \right) p \left( \lambda |  \mathcal{S} \right) \\
&=  \frac{ \lambda^{-1} \exp \left[ - \frac{1}{2 \lambda} \left( \frac{ \left( m_{i} - \lambda \right)^{2} }{\bar{w}_{i} } - \frac{ \left( n_{i} - \lambda \right)^{2} }{\bar{v}_{i}} \right) \right] }{\int_{\alpha}^{\beta} \mathrm{d} \lambda \,  \lambda^{-1} \exp \left[ - \frac{1}{2 \lambda} \left( \frac{ \left( m_{i} - \lambda \right)^{2} }{\bar{w}_{i} } - \frac{ \left( n_{i} - \lambda \right)^{2} }{\bar{v}_{i} } \right) \right] }, 
\end{align*}

\noindent with the evidence

\begin{equation*}
p \left( m_{i}, n_{i} | \mathcal{S} \right) = \frac{1}{2\pi \left(\beta - \alpha \right) \sqrt{ \bar{w}_{i} \bar{v}_{i} } } \int_{\alpha}^{\beta} \mathrm{d} \lambda \,  \lambda^{-1} \exp \left[ - \frac{1}{2 \lambda} \left( \frac{ \left( m_{i} - \lambda \right)^{2} }{\bar{w}_{i} } - \frac{ \left( n_{i} - \lambda \right)^{2} }{\bar{v}_{i} } \right) \right]
\end{equation*}

\subsubsection{The Two Source Model Evidence}

With two distinct sources the joint likelihood factors into

\begin{align*}
p \left( m_{i}, n_{i} | \lambda, \bar{ \mathcal{S} } \right) &= p \left( m_{i} | \lambda  \right) \cdot p \left( n_{i} | \mu  \right) \\
 &=\frac{1}{\sqrt{ 2 \pi \bar{w}_{i} \lambda } } \exp \left[ - \frac{1}{2} \frac{ \left( m_{i} - \lambda \right)^{2} }{\bar{w}_{i} \lambda} \right] \cdot \frac{1}{\sqrt{ 2 \pi \bar{v}_{i} \mu } } \exp \left[ - \frac{1}{2} \frac{ \left( n_{i} - \mu \right)^{2} }{\bar{v}_{i} \mu} \right] \\
 &=\frac{1}{2 \pi \sqrt{ \bar{w}_{i} \bar{v}_{i} } } \, \lambda^{-\frac{1}{2}} \exp \left[ - \frac{1}{2} \frac{ \left( m_{i} - \lambda \right)^{2} }{\bar{w}_{i} \lambda} \right] \cdot \mu^{-\frac{1}{2}} \exp \left[ - \frac{1}{2} \frac{ \left( n_{i} - \mu \right)^{2} }{\bar{v}_{i} \mu} \right] .
\end{align*}

Taking two independent priors,

\begin{align*}
p \left( \lambda, \mu | \bar{ \mathcal{S} } \right) &= p \left( \lambda | \bar{ \mathcal{S} } \right) \cdot p \left( \mu | \bar{ \mathcal{S} } \right) \\
 &= \left\{ \begin{array}{rc}  \left( \delta - \gamma  \right)^{1} \cdot \left( \zeta - \epsilon \right)^{-1} ,& \gamma \leq \lambda \leq \delta , \epsilon \leq \mu \leq \zeta \\ 0 ,& \mathrm{else} \\ \end{array} \right. ,
\end{align*}

\noindent gives the posterior

\begin{align*}
p \left( \lambda, \mu | m_{i}, n_{i}, \bar{ \mathcal{S} } \right) &= p \left( \lambda | m_{i}, n_{i}, \bar{ \mathcal{S} } \right) \cdot p \left( \mu | m_{i}, n_{i}, \bar{ \mathcal{S} } \right) \\
 &\propto p \left( m_{i}, n_{i} | \lambda, \bar{ \mathcal{S} } \right) p \left( \lambda | \bar{ \mathcal{S} } \right) \cdot p \left( m_{i}, n_{i} | \mu, \bar{ \mathcal{S} } \right) p \left( \mu | \bar{ \mathcal{S} } \right) \\
 &= \frac{ \lambda^{-\frac{1}{2}} \exp \left[ - \frac{1}{2} \frac{ \left( m_{i} - \lambda \right)^{2} }{\bar{w}_{i} \lambda} \right] }{ \int_{\gamma}^{\delta} \mathrm{d} \lambda \, \lambda^{-\frac{1}{2}} \exp \left[ - \frac{1}{2} \frac{ \left( m_{i} - \lambda \right)^{2} }{\bar{w}_{i} \lambda} \right] } \frac{ \mu^{-\frac{1}{2}} \exp \left[ - \frac{1}{2} \frac{ \left( n_{i} - \mu \right)^{2} }{\bar{v}_{i} \mu} \right] }{ \int_{\epsilon}^{\zeta} \mathrm{d} \mu \, \mu^{-\frac{1}{2}} \exp \left[ - \frac{1}{2} \frac{ \left( n_{i} - \mu \right)^{2} }{\bar{v}_{i} \mu} \right] } ,
\end{align*}

\noindent along with the evidence

\begin{align*}
p \left( m_{i}, n_{i} | \bar{ \mathcal{S} } \right) &= \frac{ p \left( m_{i}, n_{i} | \lambda, \mu,  \bar{ \mathcal{S} } \right) p \left( \lambda, \mu, \bar{ \mathcal{S} } \right) }{ p \left( \lambda, \mu | m_{i}, n_{i}, \bar{ \mathcal{S} } \right) }\\
 &= \frac{ p \left( m_{i}, n_{i} | \lambda, \bar{ \mathcal{S} } \right) p \left( \lambda | \bar{ \mathcal{S} } \right) }{ p \left( \lambda | m_{i}, n_{i}, \bar{ \mathcal{S} } \right) } \cdot \frac{ p \left( m_{i}, n_{i} | \mu, \bar{ \mathcal{S} } \right) p \left( \mu | \bar{ \mathcal{S} } \right) }{ p \left( \mu | m_{i}, n_{i}, \bar{ \mathcal{S} } \right) } \\
&= \frac{ \left( \delta - \gamma  \right)^{1} \cdot \left( \zeta - \epsilon \right)^{-1} }{ 2 \pi \sqrt{ \bar{w}_{i} \bar{v}_{i} } } \int_{\gamma}^{\delta} \mathrm{d} \lambda \, \lambda^{-\frac{1}{2}} \exp \left[ - \frac{1}{2} \frac{ \left( m_{i} - \lambda \right)^{2} }{\bar{w}_{i} \lambda} \right] \cdot \int_{\epsilon}^{\zeta} \mathrm{d} \mu \, \mu^{-\frac{1}{2}} \exp \left[ - \frac{1}{2} \frac{ \left( n_{i} - \mu \right)^{2} }{\bar{v}_{i} \mu} \right]
\end{align*}

\pagebreak

\section{Examples} 

\begin{figure}[h!]
\centering
\includegraphics[width=4.0in]{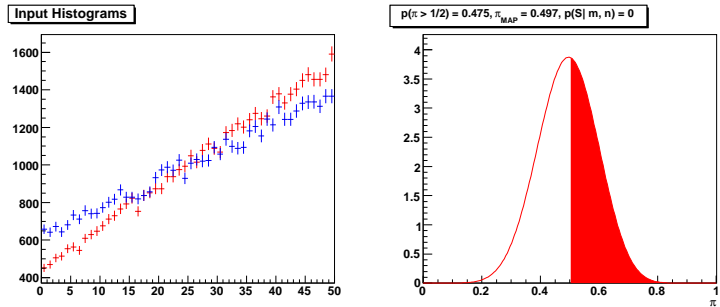}
\caption{Comparison of two histograms drawn from different distributions.
\label{fig:dataDataComp}}
\end{figure}

\begin{figure}[h!]
\centering
\includegraphics[width=4.0in]{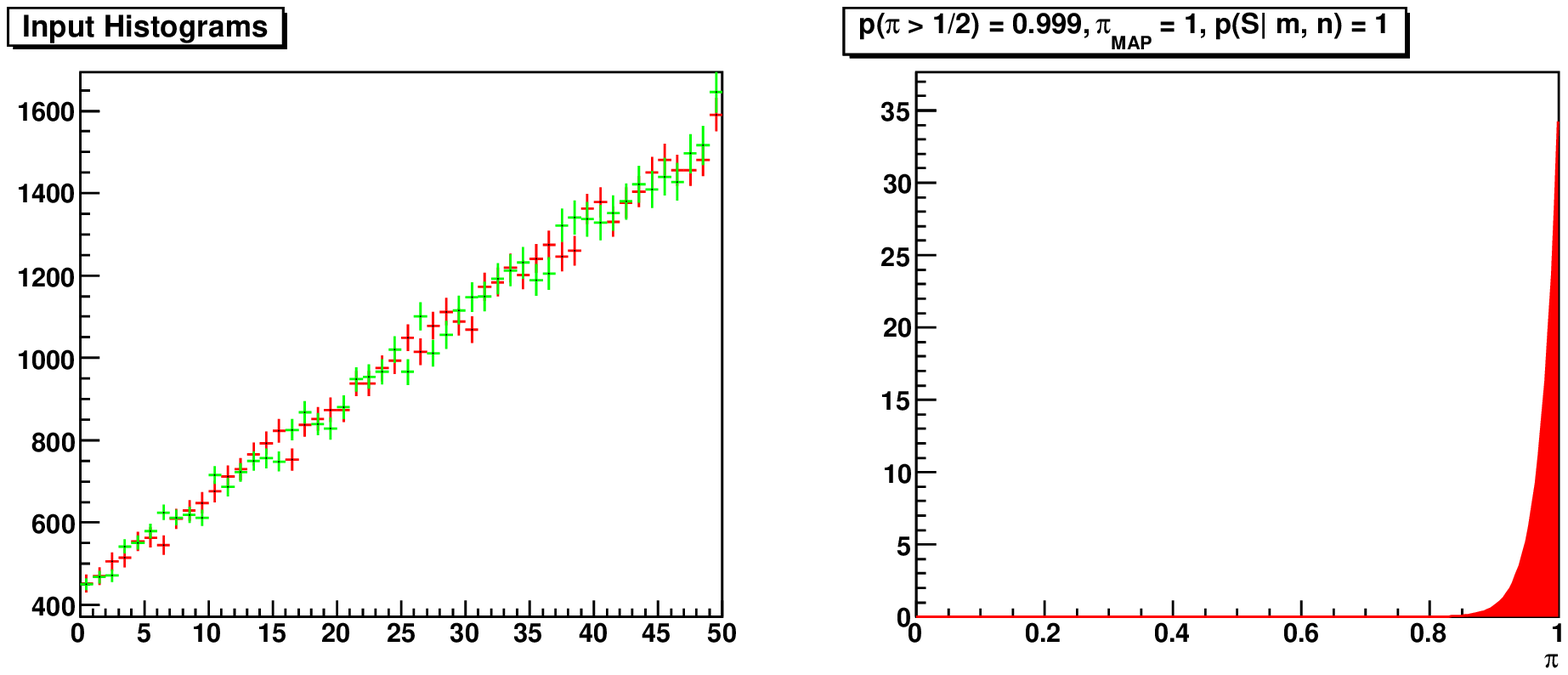}
\caption{Comparison of two histograms drawn from the same distribution, with the red sample generated from importance sampling.
\label{fig:dataSimuComp}}
\end{figure}

\begin{figure}[h!]
\centering
\includegraphics[width=4.0in]{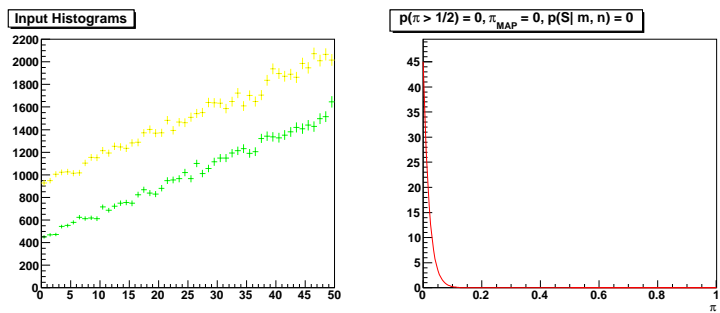}
\caption{Comparison of two histograms drawn from the same distribution, with both samples generated from importance sampling.
\label{fig:simuSimuComp}}
\end{figure}

\pagebreak

\section{Model Ensemble Tests}

The capability of the Bayesian approach is evident when considering an ensemble of histogram comparisons models.  

At each iteration in the ensemble a true model is randomly selected between $\mathcal{S}$ and $\bar{\mathcal{S}}$.  In the former case, the two histograms to be compared are drawn from a randomly generated multinomial distribution with 50 bins,

\begin{equation*}
p_{i} = \frac{ \tilde{p}_{i} }{ \sum_{j} \tilde{p}_{j} },
\end{equation*}

\begin{equation*}
\tilde{p}_{i} \sim \mathrm{U} \left( 0, 1 \right).
\end{equation*}

The latter case samples one histogram from the randomly generated multinomial distribution, but draws the second from a perturbed distribution with

\begin{equation*}
q_{i} = \frac{ \tilde{q}_{i} }{ \sum_{j} \tilde{q}_{j} } ,
\end{equation*}

\begin{equation*}
 \tilde{q}_{i} = \left(1 + \frac{2}{5} \cdot \left( r_{i} - \frac{1}{2} \right) \right) p_{i} ,
 \end{equation*}
 
\begin{equation*}
 r_{i} \sim \mathrm{U} \left( 0, 1 \right).
\end{equation*}

\noindent The two distributions are kept similar in order to assess the performance of each algorithm when comparison is highly nontrivial.

The two histograms generated for each model are then evaluated with three tests:

\vspace{2mm}
\begin{centering}

\begin{tabular}{l l}
\textbf{Non-Mixture Posterior} &\textbf{:} $p \left( S | \mathbf{m}, \mathbf{n} \right) > 0.5$ \vspace{1mm}\\
\textbf{Mixture MAP} &\textbf{:} $\pi_{\mathrm{MAP}} > 0.97$ \vspace{1mm} \\
\textbf{Kolmogorov-Smirnov} &\textbf{:} A 95\% significance KS test \\
& \hspace{2mm} implemented in the ROOT \cite{Root1997} physics library \\
\end{tabular}

\end{centering}
\vspace{2mm}

\begin{figure}[h!]
\centering
\subfigure[]{\includegraphics[width=4.0in]{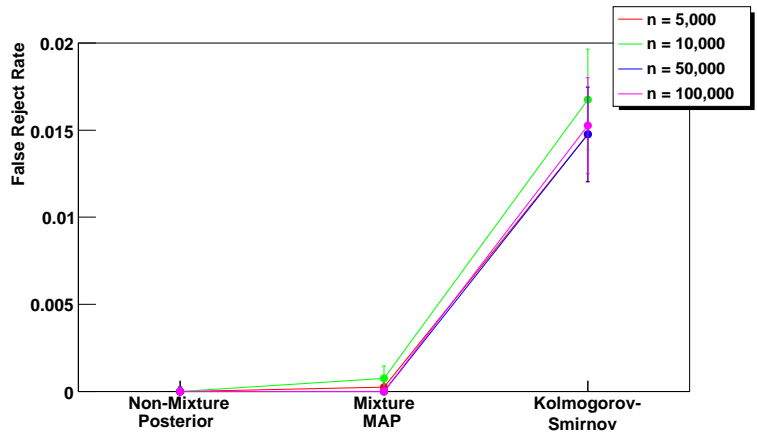}}
\subfigure[]{\includegraphics[width=4.0in]{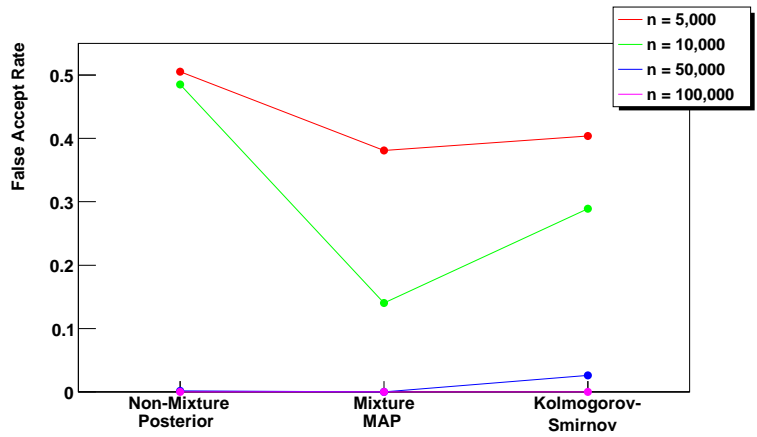}}
\caption{ (a) False reject and false accept rate for each comparison algorithm.
\label{fig:modelEnsembleTest}}
\end{figure}

\noindent False acceptances and false rejections were tabulated for four ensembles, each with 2000 independent models but different total bin contents $n$.  The respective rate posteriors were calculated assuming a binomial likelihood and beta prior, $\mathrm{Be} \left( 0.5, 0.5 \right)$; the posterior modes and 68.3\% confidence intervals for each algorithm are plotted in Figure \ref{fig:modelEnsembleTest}.

Note that in the two cases with the smallest statistics, $n = 5,000$ and $n = 10,000$ the variance of the bin contents equals or surpasses the expected difference between the two underlying distributions and the large false accept rates are to be expected.

In all cases the Bayesian mixture test, $\pi_{\mathrm{MAP}} > 0.97$, is superior to KS.  The non-mixture test outperforms KS in almost all cases as well, failing only in the small statistics ensembles where small discrepancies tend to be washed out by the bin content variances.

\pagebreak

\pagebreak

\section{Conclusion}

A Bayesian mixture model has been developed to test whether two histograms are consistent, in that they are more likely to have been drawn from a single distribution rather than two distinct distributions.  The model is extended to handle histograms generated from importance sampling, resulting in a robust and powerful approach to the comparison of histograms populated by both data and simulation.  Said power is evident with studies of a large model ensemble.

A C++ implementation of the Bayesian mixture model utilizing the ROOT \cite{Root1997} data analysis framework is available at \url{http://web.mit.edu/~betan/www/code.html}.

\section{Acknowledgements}

I thank Chris Jones, Steve Voinea, and Matt Walker for helpful discussion and comments.

\end{document}